\documentclass{iaus}
\usepackage{graphicx}

\title[Instabilities in GRB central engine] 
{Instabilities in the Gamma Ray Burst central engine. What makes the jet 
variable?}

\author[Agnieszka Janiuk et al.]   
{Agnieszka Janiuk$^1$,  Ye-Fei Yuan$^2$,
Rosalba Perna$^3$ \and Tiziana Di Matteo$^4$}

\affiliation{$^1$Center for Theoretical Physics, Polish Academy of Sciences, Al. Lotnikow 32/46, 02-668 Warsaw, Poland \\[\affilskip]
$^2$ Department of Astronomy, University of Science and Technology of China, Chinese Academy of Sciences, Hefei, Anhui 230026, P.R. China \\[\affilskip]
$^3$ JILA and Department of Astrophysical and Planetary Sciences,
University of Colorado,
Boulder, CO 80309 USA \\[\affilskip]
$^4$ Physics Department, Carnegie Mellon University, 5000 Forbes Avenue, Pittsburgh, PA 15232  \\[\affilskip]
}

\pubyear{2010}
\volume{275}  
\pagerange{1-1}
\jname{Jets at all scales}
\editors{Gustavo E. Romero, Rashid A. Sunyaev \& Tomaso Belloni}
\begin{document}

\maketitle

\begin{abstract}
Both types of long and short gamma ray bursts involve a stage of a 
hyper-Eddington accretion of hot and dense plasma torus onto a newly born 
black hole.
The prompt gamma ray emission originates in jets at some distance from this 
'central engine' and in most events is rapidly variable, having a form 
of sipkes and subpulses. This indicates at the variable 
nature of the engine itself, for which a plausible mechanism is an 
internal instability in the accreting flow.
We solve numerically the structure and evolution of 
the neutrino-cooled torus. 
We take into account the detailed treatment of the microphysics in the 
nuclear equation of state that includes the neutrino trapping effect.
The models are calculated
for both Schwarzschild and Kerr black holes.
We find that for sufficiently large accretion rates ($>\sim 10 M_{\odot}$ s$^{-1}$
for non-rotating black hole, 
and $>\sim 1 M_{\odot}$ s$^{-1}$ for rotating black hole, depending on its spin), 
the inner regions of the disk become opaque, while the helium nuclei are being 
photodissociated. The sudden change of pressure in this region leads to
the development of a viscous and thermal instability, and the
neutrino pressure acts similarly to the radiation pressure 
in sub-Eddington disks.
In the case of rapidly rotating black holes, the instability
 is enhanced and appears for much lower accretion rates. 
We also find the important and possibly further destabilizing
 role of the energy transfer from the rotating 
black hole to the torus via the magnetic coupling.
\end{abstract}

\section{Introduction}

Roughly 80 per cent of the observed gamma ray bursts exhibit a substructure in
their time profiles and the timescales of the sub-pulses are about $10^{3}$-$10^{4}$
times shorter than the total duration of the event, estimated by $T_{90}$ 
(see \cite{piran05} for a review).
This effect is most probably connected with the rapidly variable conditions
 within the jet plasma, such as the Lorentz factor. Therefore, since 
the energy 
input into the jet is required to vary, the activity of the central engine that 
produces this energy should not be stationary but rather
 change in much shorter timescale than the engine lifetime.

The central engine of the gamma ray burst is presumably a newly born black hole, 
surrounded by an accretion disk built from either a remnant of the disrupted 
companion star in the compact binary system, or the fallback material from the 
hypernova envelope.
Such disk is extremely hot and dense due to the large accretion rate which can be
 on the order of 0.1 up to a few solar masses per second. Such physical conditios
make the plasma totally opaque to photons and partially opaque to neutrinos 
(\cite{dpn02}), which are produced in the
 nuclear reactions and provide a source of cooling to the disk (\cite{popham99}).

The gravitational instability of the outer parts of the hyperaccreting disk in the GRB 
central engine was suggested first by \cite{perna} and studied in the numerical model
of \cite{chen07}. Such instability may be a source for the late time activity and 
longer duration flares from the central engine.
The prompt emission however is more plausibly explained with the instability 
discussed in \cite{jan07}, and more recently in \cite{jan10}, which arises in the 
regions of the disk closest to the 
central black hole.

\section{Instability mechanism}

The time dependent model of the hyperaccreting disk in the GRB central engine,
presented in \cite{jan04}, was based on a simplified equation of state, where the 
total pressure was given by a sum of the ideal gas, radiation 
and degenerate electron pressure. The more elaborate EOS, used in \cite{jan07},
invokes self-consistently a chemical equilibrium of species such as protons, neutrons, 
electron-positron pairs and alpha particles. The reaction rates are 
computed under the assumption of beta equilibrium and charge neutrality, and the 
neutrino opacities are due to their absorption and scattering. To the total pressure 
contribute both partially trapped neutrinos and alpha particles, which are continuously
photodissociated and recombinated in the plasma.

\begin{figure}[b]
 \includegraphics[width=2.7in]{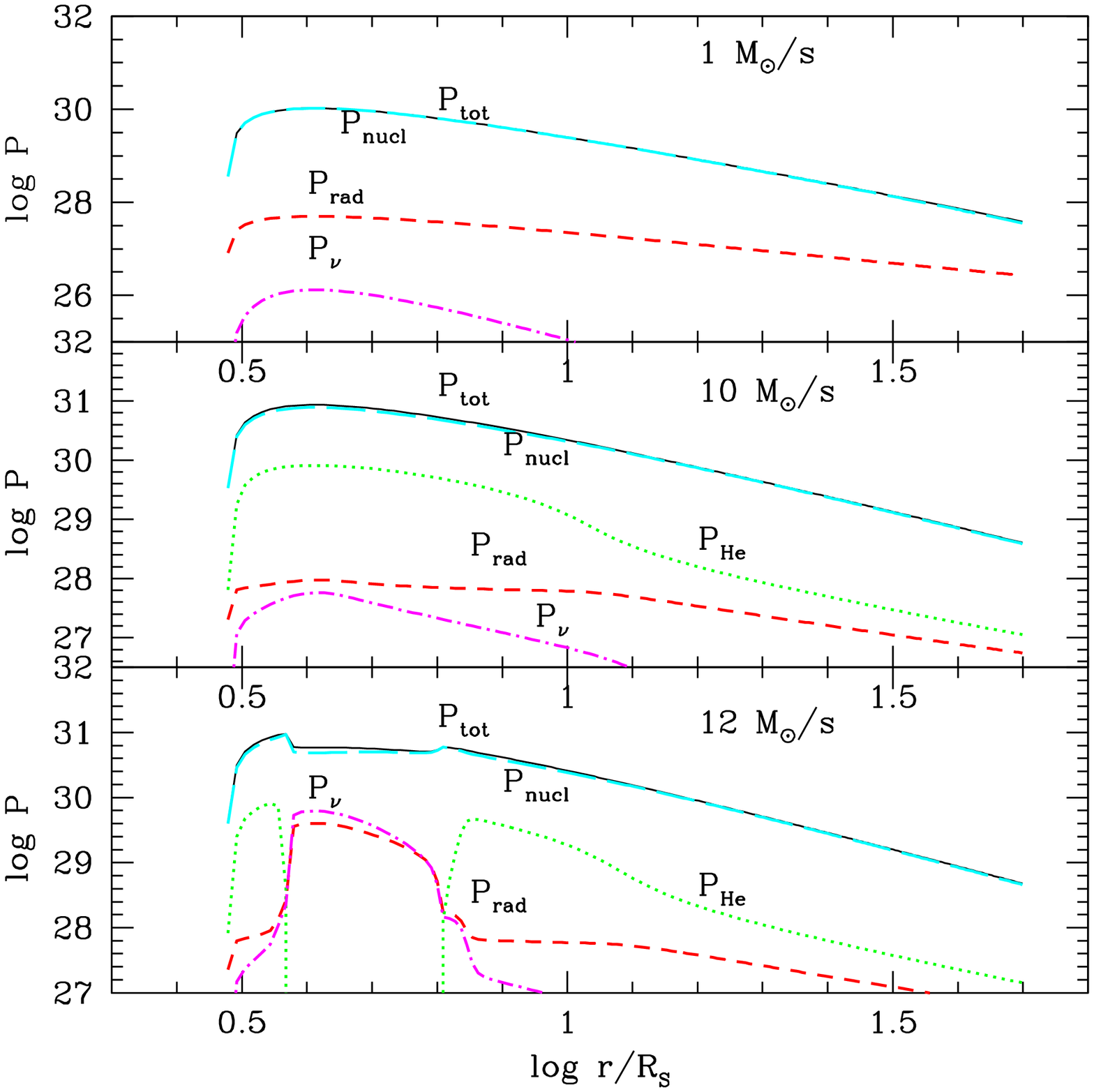} 
 \includegraphics[width=2.7in]{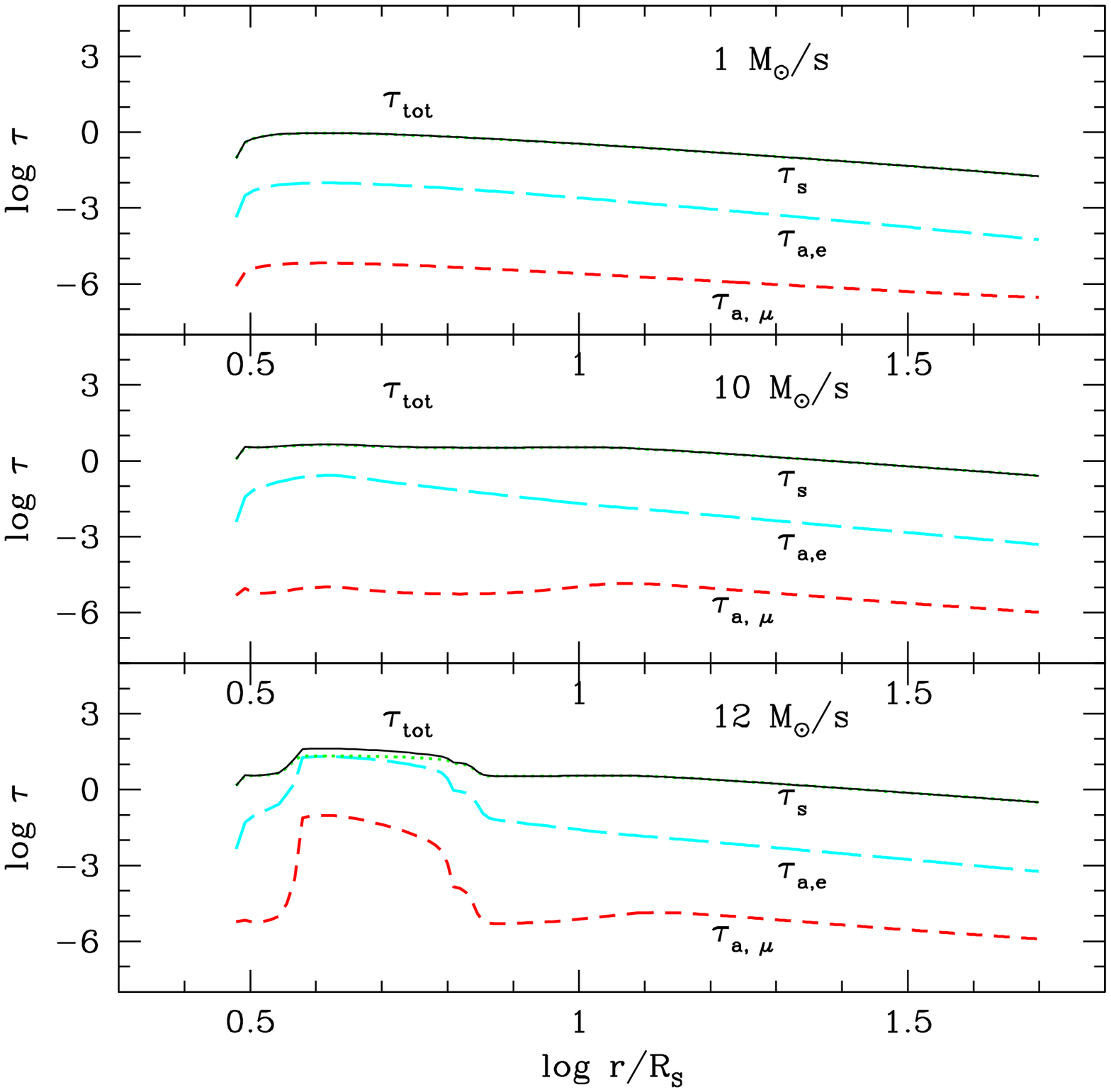} 
 \caption{Radial profiles of the total pressure and its 
components: nuclear, helium, neutrino and 
radiation pressure (left),  and neutrino total opacity and its components due to 
absorption, for muon and electron neutrinos, and scattering (right).
The models were computed for a non-rotating black hole, 
with three values of accretion rate, as indicated in the 
panels. The black hole mass is 3 $M_{\odot}$ and viscosity $\alpha=0.1$}
   \label{fig:fig1}
\end{figure}

As results from our first calculations, at very high accretion rates, which 
for a non-rotating black hole should be at least 10 $M_{\odot}$ s$^{-1}$, a thermal
instability arises in the innermost disk region close to the inner edge. 
This instability
is associated with helium photodisintegration, which was found to lead
to the local accretion rate fluctuations already by e.g. \cite{mfw99}. 
What can be seen from our Fig. \ref{fig:fig1}, bottom panel, the helium 
photodissociacion in a narrow strip of the accreting disk leads to a phase 
transition in the nuclear material. Due to this transition, the vanishing 
helium pressure is compensated by the neutrino pressure. The plasma
becomes opaque to neutrinos and the neutrino absorption optical depths
increase locally for both muon and electron flavor species. 
However, the net total pressure drops somewhat in the unsstable region
and the resulting density profile in the disk is inversed. The unstable strip
is much hotter and less dense than its surroundings. This leads to a faster 
replenishing of the strip than would be implied by the constant accretion rate
in a stable situation.
The accretion rate varies in both time and radius, which in turn leads to the
neutrino luminosity variations and variable energy output from the disk.

The instability mechanism is much similar to the well known radiation pressure 
instability, operating in the accretion disks of microquasars and AGN 
(e.g., \cite{jan02}). In case of gamma ray burst disks, the role of radiation
is taken by the neutrino pressure, which has a similar dependency on temperature.
However, the mechanism stabilizing the disk in its hot phase, which in the radiation 
pressure dominated disk is due to advection, is not sufficient in the neutrino cooled 
disks.

\section{Effect of black hole rotation and its energy extraction via magnetic fields}

\begin{figure}[b]
 \includegraphics[width=2.7in]{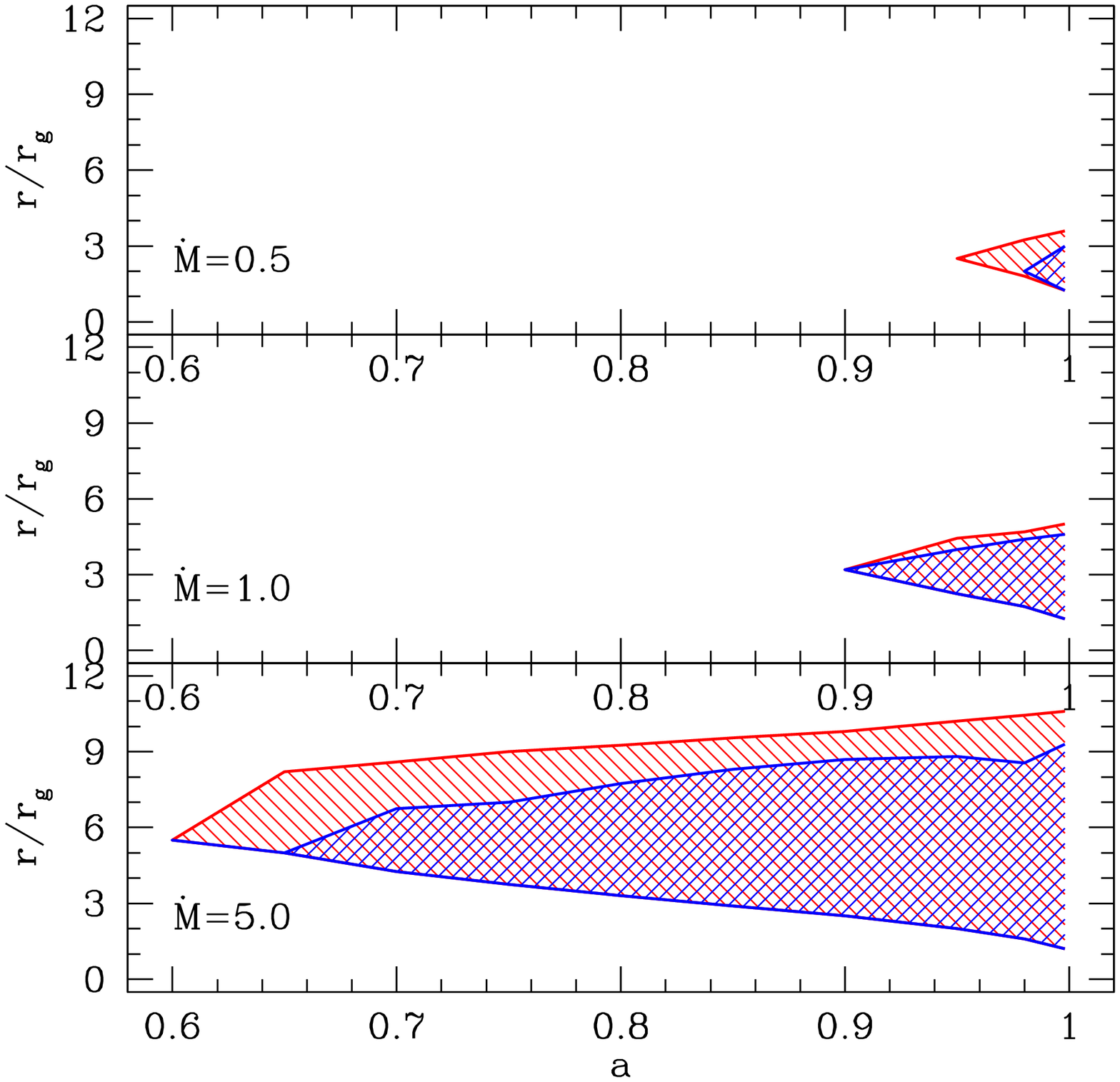} 
 \includegraphics[width=2.7in]{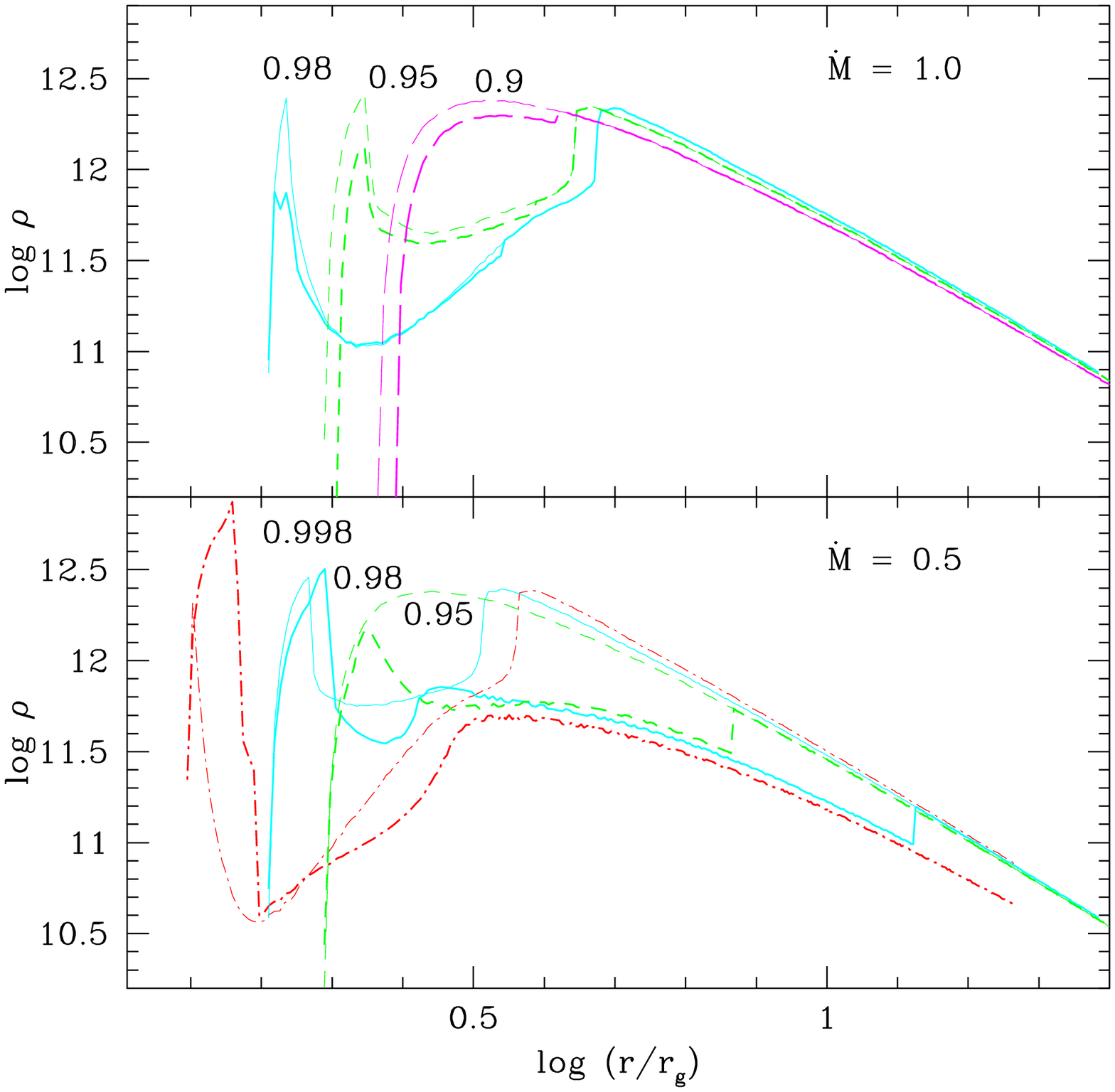} 
 \caption{Extension of the unstable zones, depending on the black hole spin (left), 
and radial density profiles (right). The values of the accretion rate are marked in 
the panels and the black hole mass is $4 M_{\odot}$. The visocsity in the left figure 
is $\alpha=0.1$ (line shades regions) or $\alpha=0.3$ (cross-shaded). The right figure 
is for
 $\alpha=0.1$, and the spin parameters are indicated at the top of each line. The 
thicker lines show the results for the disk magnetically coupled with a 
rotating black hole and the thinner lines are for neglected magnetic coupling.}
   \label{fig:fig2}
\end{figure}

The accretion rates needed for the thermal instability to appear do not have to be 
extremely large if the disk surrounds a rotating black hole.
Such a black hole would be a natural outcom e.g. of a collapsing Wolf Rayet star with
a rapid rotation and is much more plausible for powering the GRB jets than a 
Schwarzschild black hole. As estimated in \cite{jan08}, the longest duration GRBs 
require for their lifetimes large black hole spins, $a>0.9$. This is because otherwise
the realistic distributions of angular momentum within the collapsing envelope
do not provide a sufficient condition for a long duration central engine lifetime.

In Figure \ref{fig:fig2} we show that the thermal instability occurs in the
Kerr black hole disks at quite low accretion rates, e.g. at 0.5 $M_{\odot}$ s$^{-1}$, 
as is plausible for the collapsar scenario, or a few $M_{\odot}$ s$^{-1}$, as 
more plausible in case of merging neutron stars. This result slightly depends on 
the adopted disk viscosity ($\alpha$ parameter) and is very sensitive to the
black hole spin.

The additional effect which needs to be taken into account in case of the Kerr
black hole, is the transfer of its rotational energy to the disk and vice-versa.
In other words, the black hole can be spun up by accretion, or spun down if the
angular velocity in the inner radii of the differentially rotating disk
exceeds that of the black hole. The energy transfer proceeds via the closed 
magnetic field
lines and leads to the additional torque due to the black hole coupling 
(\cite{wang02}).
At the same time, the open magnetic field lines can transfer the black hole rotation 
energy to the remote load and power the jet via the Blandford - Znajek process.
We found that such a coupling does not stabilize the
disk against the thermal instability discussed above. The unstable region can be 
shrunken but also shifted outwards, while the density drop in this region can even be deeper.
Another effect is, that in case of the stable disk with a moderate neutrino pressure 
and some helium nuclei, the additional heating due to the rotating black hole 
coupling may lead to another type of unstable behaviour (\cite{lei09}).

\section{Conclusions}

According to our results, the longest duration GRBs are powered by the fast 
rotation of the black hole. They sould be rapidly variable at the beginning of 
their prompt emission, due to the instabilities in the accretion process. 
Later on, the accretion rate drops and the black hole spins down, so that the
instability is no longer operating and the burst profile will smoothen.
On the other hand, also a moderately rotating black hole can be produced after the 
hypernova collapse. If the accretion rate is large, the black hole spin up 
may later allow for sufficient conditions for the instability.
The GRB observed properties may therefore serve to test the role of black hole spin
in the jet production in accreting black hole systems, other than X-ray binaries and AGN
(\cite{fender10}).

{\bf Acknowledgments} This work was supported in part by grant NN 203 512638 
from the Polish Ministry of Science.

\begin{discussion}

\discuss{Koide}{You noted the energy transport from the rotating black hole
to the disk through the closed magnetic field lines. As far as we assume ideal 
MHD or a force-free condition, it is difficult to store the energy in the disk.
I think the energy extracted from the black hole is stored in magnetic field
outside the disk, where the magnetic configuration changes. Could you comment on
this issue?}

\discuss{Janiuk}{The disk plasma is fully ionized so we assume the disk to be
perfectly conducting and the magnetic filed lines frozen in. When the 
accretion proceeds, the field lines are advected and its configuration
indeed may be changed. Our assumption is in fact a steady state approximation.} 

\end{discussion}

\end{document}